\documentclass[debug]{rmaa}

\usepackage{paralist}

\usepackage{psfrag,color}

\usepackage[latin1]{inputenc}

\usepackage{mathptmx}
\usepackage{txfonts}
\usepackage{textcomp}
\usepackage{ulem}
\usepackage{graphicx}	
\usepackage{amssymb}
\usepackage{natbib}

\def\msun{M$_{\odot}$}

\def\fs{{\sl FS\,Aur}}

\title{Revisiting {\sl FS Aurigae} and its  triple cataclysmic variable system hypothesis} 

\author{
Carlos E. Chavez,\altaffilmark{1} 
Andres Aviles,\altaffilmark{1}
Nikolaos Georgakarakos,\altaffilmark{2}
Cesar Ramos,\altaffilmark{3,5}
Hector Aceves,\altaffilmark{4}
Gagik Tovmassian,\altaffilmark{4}
Sergey Zharikov,\altaffilmark{4}
}

\altaffiltext{1}{FIME-UANL. M\'exico.}
\altaffiltext{2}{New York University Abu Dhabi, UAE.}
\altaffiltext{3}{FCFM-UANL. M\'exico}
\altaffiltext{4}{IA-UNAM, Ensenada. M\'exico }
\altaffiltext{5}{CINVESTAV, Ciudad de M\'exico. M\'exico }

\shortauthor{Chavez et al.}
\shorttitle{Revisiting FS Aurigae}

\fulladdresses{
\item Chavez C.E and Aviles A: Universidad Auton\'oma de Nuevo Le\'on, Facultad de Ingenier\'{i}a Mec\'anica y El\'ectrica, San Nicol\'as de los Garza, NL, M\'exico (Carlos.ChavezPch@uanl.edu.mx).
\item Georgakarakos N: New York University Abu Dhabi, Saadiyat Island, P.O. Box 129188, Abu Dhabi, UAE.
\item Ramos~C: Universidad Auton\'oma de Nuevo Le\'on, Facultad de Ciencias F\'{i}sico--Matem\'aticas, San Nicol\'as de los Garza, N.L. M\'exico.
\item Ramos~C: Centro de Investigaci\'on y de Estudios Avanzados del Instituto Polit\'ecnico Nacional, San Pedro Zacatenco, Ciudad de M\'exico, 07360, M\'exico.
\item Aceves H, Tovmassian G, Zharikov S: Universidad Nacional Aut\'onoma de M\'exico, Instituto de Astronom\'\i{}a, Ensenada 22860, B.C., M\'exico,. 
}

\listofauthors{C.E. Chavez et al.}
\indexauthor{Chavez, C.E.}
\indexauthor{Aviles, A.}
\indexauthor{Georgakarakos, G.}
\indexauthor{Ramos, C.}
\indexauthor{Aceves, H.}
\indexauthor{Tovmassian, G.}
\indexauthor{Zharikov, S.}

\abstract{A very long term variability (VLPP), with period of 875 days, was observed in the long-term light curve of {\sl FS Aurigae} (\fs) in 2011. This periodicity was calculated on 6 cycles. We re--examine the periodicity with new observations over of the past 5 yrs.
A total of 18 yrs of observations confirm the hypothesis of a third body perturbing in a secular way the cataclysmic variable (CV). Improvements to the model such as eccentric and inclined orbits for the third body and a binary post--Newtonian correction are considered.
We confirm the VLPP of \fs\ and find the new period of $857 \pm 78$ days.
 The secular perturbations are most efficient when the mass of the third body is M$_3 \approx$ 29 M$_J$, much less than the 50 M$_J$ reported in 2011. We estimate the effect of the third body on the mass transfer rate and the brightness of the system. We consider alternative scenarios for the VLPP. The new data and analysis supports the hypothesis that \fs~is a CV in a triple system. }

\resumen{
Una variabilidad de muy larga duraci\'on (VLPP), con un periodo de 875 d\'i{}as, fue observado en la curva de luz de \fs~en 2011. El periodo fue calculado con 6 ciclos. Reexaminamos el periodo con nuevas observaciones de los pasados 5 a\~nos. Un total de 18 a\~nos de observaciones confirman la hip\'otesis de un tercer cuerpo perturbando de manera secular la variable catacl\'i{}smica (VC). Mejoras al modelo como \'orbitas exc\'entricas e inclinadas para el tercer cuerpo y una correcci\'on  post--Newtoniana para la binaria son consideradas.
Confirmamos el VLPP de \fs~y encontramos el nuevo periodo de 857 $\pm$ 78 d\'i{}as. Las perturbaciones seculares son m\'as eficientes cuando la masa del tercer cuerpo es M$_3 \approx$ 29M$_J$, menor a M$_3 \approx$ 50M$_J$ reportado en 2011. Estimamos el efecto del tercer cuerpo en la tasa de transferencia de masa y del brillo del sistema. Consideramos otras explicaciones para el VLPP. Estos nuevos datos y an\'alisis apoyan la hip\'otesis de una VC triple para \fs.}

\addkeyword{Stars: binaries (including multiple)}
\addkeyword{Stars: individual (FS Aur)}
\addkeyword{Stars: novae, cataclysmic variables}

\begin{document}


\maketitle

\section{Introduction}
\label{sec:intro}

\fs\ is a Cataclysmic Variable (CV) that shows a wide range of light periodic signals. It has a short orbital period of just 85.7 min (Thorstensen et al. 1996), a  long photometric period of 205.5 min (Tovmassian et al. 2003) and a long spectroscopic period of 147 min (Tovmassian et al. 2007). The latter two periods are attributed to the precession of a fast rotating magnetic white dwarf and its beat with the orbital period, respectively (see Table \ref{tab:periods} for details).
All these frequencies were discussed in more detail in Chavez et al. (2012, hereafter CH2012). In that paper we showed the presence of a very long photometric period (VLPP) modulation observed in the long-term \fs\ light curve, with a 2--mag amplitude and a period around 900 days. We argued that the origin of such modulation could be a third substellar-body (25 to 65 times Jupiter\textquoteright s~mass) that perturbs the eccentricity of the inner binary star system. 

This triple--system hypothesis provided an explanation for the VLPP, and it was also suggested that it might give a plausible answer for other observed peculiarities of \fs. More importantly is perhaps  the fact that it offers a new possibility for detecting planets in accretion disk environments, where other methods fail.

There are other binary systems claimed to have a third object in a close orbit. {\sl LX Ser} possess an extra component of 7.5 times the mass of Jupiter that explains a sinusoidal oscillation observed in the O -- C diagram with a period of 22.8 years (Li et al. 2016). Another example is {\sl V893 Scorpi} where  observed variations of the eclipse period of 10.2 years are interpreted as a light travel time effect caused by the presence of a giant planet with 9.5 times the mass of Jupiter (Bruch 2014). Finally {\sl DP Leonis} (Beuermann et al. 2011), {\sl HW Vir}  (Lee et al. 2009), {\sl NN Ser} (Beuermann et al. 2010), {\sl NY Virginis} (Qian et al. 2012a), RR Caeli (Qian et al. 2012b) and KIC 5095269 (Getley et al. 2017) are part of this small group of post-CE binaries suspected to possess planets. 

The purpose of this paper is to make use of 5 more years of observations of FS Aurigae to see whether the VLPP signal reported in CH2012 is stronger or, on the contrary, is disappearing. We also want to model the hierarchical triple hypothesis in a more realistic manner by including eccentric and inclined orbits and also first order post--Newtonian correction, that is a first order general relativity correction. Then studying the effect these complications have on the range of possible values on mass and semi--major axis that may explain the VLPP by secular perturbations on the Cataclysmic Variable.
This paper is organized as follows.
In \S\ref{Sec:observations}, we review observational data of \fs\ in search of the very long photometric period (VLPP).  
In \S3.1 we revisit the initial conditions used in our previous research and more recent and more accurate parameters for our system are indicated.
In \S3.2 we examine the scenario where the perturber moves on a circular and coplanar orbit, whose period is much shorter than the long period, and yet produces a binary eccentricity variation with the latter period by secular perturbations. In \S3.3 we extend this to eccentric and inclined orbits. The range and properties of the allowed solutions are shown. In \S3.4 we check if the VLPP could be explained as a consequence of the precession effect of the orbit due to first order general relativity corrections. In \S3.5 we make an order of magnitude estimation for the mass transfer rate and the brightness of the system. In \S4 we explore alternative scenarios for the explanation of the observed VLPP, with particular attention to the cyclic magnetic variation. In \S\ref{Sec:final} we provide some final comments on the new results and its observational imprint on {\sl FS Aurigae}'s features.


\section{The long and permanent photometric  behaviour of {\sl FS Aur.}}
\label{Sec:observations}

\begin{figure}[t]
\includegraphics[width=\columnwidth]{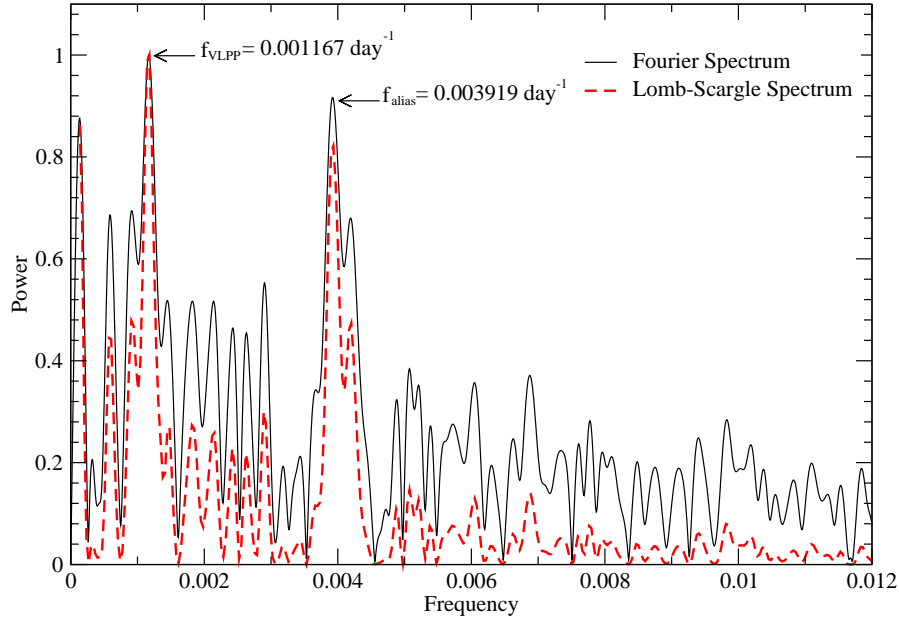}
\caption{Normalized power spectrum of the quiescent light curve of \fs\,. Solid black curve corresponds to our Fourier analysis and red dashed curve corresponds to the Lomb--Scargle method. The strongest  peak f$_{\mathrm {VLPP}}=0.001167$\,day$^{-1}$\  corresponds to the Very Long Photometric Period. The second--highest peak frequency in the power spectrum f$_{\mathrm {alias}}=0.003919$\,day$^{-1}$\  corresponds to an alias created by yearly observational cycle f$_{\mathrm {Y}} = 0.002739$\,day$^{-1}$\ and f$_{\mathrm {VLPP}}$.}
\label{fig1new}
\end{figure}

Here, we use a data set 1.4 times larger than the one used  earlier, covering more than 7,500 days of observations, coming from the AAVSO public data base. From our analysis, we conclude that the long period is still present in the light curve and confirm the phenomenon reported in CH2012.  The power spectrum of the data is displayed  in  Figure\,\ref{fig1new}. The data set spans over 20 years and  almost nine periods of $\sim 850$ days, peaking in the periodogram at the 0.001167 day$^{-1}$ frequency. The other low--frequency peak of similar strength at $f=0.003919\, {\mathrm {day}}^{-1}$ is an alias related to the one year observational cycle.
When taking into account a larger set of data, the estimated period is $857\pm 78$ days, and coincides well with the one previously reported ($875\pm 50$ days) within the estimated error.

\begin{figure}[t]
\includegraphics[width=\columnwidth]{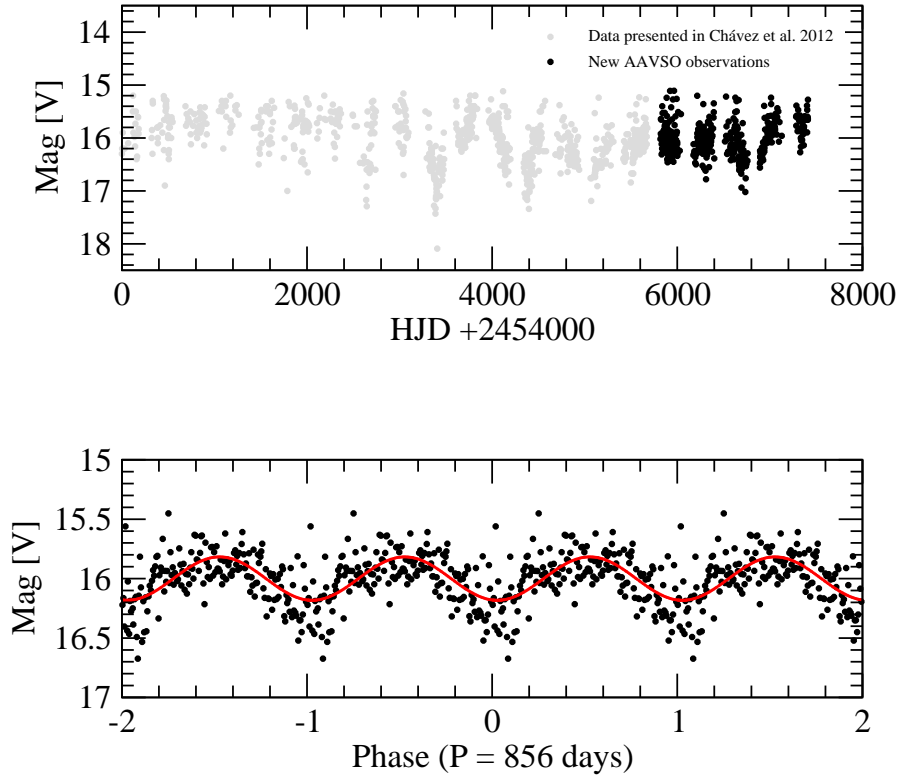}
\caption{Upper panel, long-term light curve of \fs\ over the past 20 years, 1.4 times larger than in CH2012 (black filled circles correspond to new observations). Bottom panel, folded light curve in quiescence using the VLPP of 857 days. We also show in red the best sinusoidal fit for this curve.}
\label{fig2new}
\end{figure}

The upper panel of Figure~\ref{fig2new} corresponds to the long-term light curve for \fs\ in the ${\it V}$ band.  The bottom panel of Figure~\ref{fig2new} displays the folded light curve adjusted with a VLPP period of 857 days. The amount of data for the folded light curve was reduced averaging the magnitude per phase to appreciate in detail the sinusoidal behaviour. We calculated the best sinusoidal fit for the bottom panel of Figure~\ref{fig2new}, shown in red in the figure, we found that the amplitude of the best fit is $\approx 0.4$ magnitudes, but it is also clear that the data points are disperse, then we also calculated the difference between the maximum and minimum magnitude of the observed data finding 1.1 magnitude.


\begin{table*}
\caption{Summary of periodicities detected in \fs.}

\begin{tabular}{@{}lccll}
\hline
 Name/Acronym & Value & Source & Reference & Comments \\
\hline
 Spin Period of WD (SP)                       &   $1\fm68 - 1\fm75 $ & light curve  & (Neustroev et al. 2005) & inconclusive\\
 Orbital Period   (OP)      & $85\fm79736$    & RV$^{a}$  & (Thorstensen et al. 1996) & firmly \\
 &$\pm0.00004$&core of H lines & unpublished data &established \\ 
 Long Spectr.  Period  (LSP)     & $147^m$     & RV$^{b}$     & (Tovmassian et al. 2003)  &  beat  between \\
          &     &  wings of H lines   &   & OP and LPP \\ 
 Long Phot. Period  (LPP)         & $205\fm45013$ & light curve         & (Tovmassian et al. 2003)  & stable \\
  &$\pm0.0006$& & & over $\sim3\,000^{\mathrm d}$ \\  
 $\bf{Previous}$ Very Long Phot. Period & $875^{\mathrm d}\pm50^{\mathrm d}$      & light curve        & (Chavez et al. 2012) &  based on \\ 
   (VLPP)  &         &                                      &                                           &        $\sim5\,000^{\mathrm d}$ coverage  \\
$\bf{New}$ Very Long Phot. Period & $857^{\mathrm d}\pm78^{\mathrm d}$      & light curve        & this work &  based on \\ 
   (VLPP)  &         &                                      &                                           &        $\sim7\,500^{\mathrm d}$ coverage  \\

  \bottomrule
\end{tabular} \\
$^{a}${measured in the core of emission lines} \\ 
$^{b}${measured in the extreme wings of emission lines} \\
\label{tab:periods}
\end{table*}

\section{Revisiting the triple cataclysmic variable system hypothesis}
\label{Sec:3}

A CV is a binary system that is composed by a primary  massive star, a white dwarf, and a low mass main sequence K--L type star with a predominant population of M (red dwarf) stars. They are so close to each other that the secondary star fills its Roche lobe and its surface is close to the $L_{1}$ Lagrangian point.
 
The material that the secondary loses cannot fall directly to the primary, but instead it forms a disk of material around the primary and references therein (Ritter 2008). This disk is so bright that outshines the brightness of both stars. In fact, its brightness is proportional to the mass transfer rate or to the mass accretion rate (Warner 1995). 
Therefore, if there is a change in the mass transfer rate, there will be also a change in the system's  brightness.Therefore, any change in the location of the Lagrangian $L_{1}$ point will change the mass transfer and therefore the brightness of the system.

We recall that there is a huge disparity between the VLPP and all other periods. This lead CH2012 to seek the cause of the variable mass transfer rate and therefore of the disk brightness not related to the binary itself but to propose a third body orbiting the binary.

The presence of a third body in the system would result 
in perturbing the orbit of the stellar binary on different timescales. These timescales depend on the mass, eccentricity  and semi-major axis of the unseen companion.  Therefore, knowing the period of the long--term variability of the light curve of \fs\ can help us place constraints on the
mass and  orbital configuration of the potential companion. 

For that purpose, we can make use of some previously derived analytical results.  In a series of papers, Georgakarakos (2002, 2003, 2004, 2006, 2009, 2015, 2016) studied the orbital evolution of hierarchical triple systems. Part of those studies were on the secular evolution of such systems. The analytical results derived there can give us an estimate about the frequency and the period of motion of the stellar binary. Therefore, we can estimate which mass values and orbital configurations of a hypothetical third companion can yield the secular period observed in the light curve of \fs.

For a coplanar system with a perturber on a low eccentricity orbit we make use of the results of Georgakarakos (2009),  while for coplanar systems with eccentric perturbers those of Georgakarakos (2003).  Finally,  for systems with low eccentricity orbits and low mutual inclinations ($ i < 39.23^{\circ}$,that angle is the limit before Kozai resonances becomes important as explained in Kozai (1962) ) we can use the relevant material of Georgakarakos (2004).

\subsection{Initial parameters}

Here we discuss briefly the origin of all parameters used in this work. In CH2012 we used the following parameters: total mass $M_{T} = M_{1}+M_{2}=0.84$\,\msun\, with  the primary mass $M_{1}=0.75$\,\msun, and the secondary one $M_{2}=0.09$\,\msun.

We decided to revisit these parameters, starting with the mass and radius of the secondary. Here we use the values that appear in Knigge et al. (2011), in which they use the eclipsing CVs and theoretical constrains to obtain a semi--empirical donor sequence for CVs with orbital periods  $P_{orb}\leq6h$. They give all key physical and photometric parameters of CVs secondaries, as well as their spectral types, as a function of $P_{orb}$.

We use the data that appear on the above authors' Table~6 and Table~8 to obtain the best physical parameters for \fs.
 We interpolate between values to find the best possible ones for our dynamical study, these are shown in Table \ref{tab:initial}. We obtain the following mass ratio between secondary and primary $q=M_2/M_1 = 0.1$  as shown in Table~\ref{tab:initial}.  The primary mass was obtained from Knigge et al. (2011) and is based on the value that they previously obtained in Knigge et al. (2006). That value was calculated as the mean value of the WD mass among the eclipsing CV sample available at the time $ \langle M_{1} \rangle = 0.75 \pm 0.05$\msun .  They stated that when adding new data the mean increases but not significantly, so they decided to retain the    
$ M_{1}  = 0.75$\msun~value as a representative WD mass.

\begin{table*}
\caption{Initial parameters of \fs.  }
\begin{tabular}{@{}lcl}
\hline
 Parameter & Value & Reference \\
 \hline
 Orbital Period  & 1.42996 hours  &  Thorstensen~et~al.~(1996)\\
 Orbital semi--major axis of the Binary & 0.6$R_{\odot}$ & Knigge~et~al.~(2011) \\
 Secondary star mass & 0.08 $M_{\odot}$  & Knigge~et~al.~(2011)\\
 Secondary star radius & $0.12$ $R_{\odot}$ & Knigge~et~al.~(2011) \\
 Primary star mass & 0.75 $M_{\odot}$ & Knigge~et~al.~(2011)  \\
Primary star  radius & 0.01 $R_{\odot}$ & Knigge~et~al.~(2011)  \\
 Log Secondary star mass loss rate & -10.25  \bigg(${M_{\odot} \over  yr}$\bigg) & Knigge~et~al.~(2011)\\
 Secondary star Temperature/Spectral Type & 2600/M7.0 & Knigge~et~al.~(2011)\\
 Mass ratio & 0.1 & -- \\
 \hline
\end{tabular} \\
\label{tab:initial}
\end{table*}

We performed simulations of the CV with a hypothetical third body.  In all numerical integrations, except the ones that are stated otherwise, in the subsequent subsections, we  used the high--order Runge--Kutta--Nystr\"om RKN 12(10) 17M integrator of Brankin~et~al.~(1989)  for the equations of motion  of the full three--body problem in the barycentre inertial reference frame. In our integrations, the total energy is monitored and it is conserved up to $10^{-5}$, or better, for all experiments. At each time step, the instantaneous eccentricity of the binary is computed. 

As pointed out in CH2012, tidal deformation of the stars  in the close binary  three-body problem can be an important effect. However, CH2012 have shown that these tidal effects are not important for this system and the two objects can be considered as point masses.

\subsection{The third body on a close near-circular and co--planar orbit}

Hierarchical triple systems consist of two stars in a close orbit and a third body orbiting the barycentre of the close binary.

In Chavez et al. 2012 we ruled out that the VLPP could correspond directly to the period of a third body, since the object would be too far for having an important effect on the inner binary. There, we performed a series of numerical integrations in which we proved that indeed the effect is very small and could not explain the VLPP of the CV. Instead, we concluded that a third light-weight body can produce a disturbance on the central binary and such perturbation may have a much longer period compared to the orbital period of the perturber (e.g. Mazeh \& Shaham~1979, Soderhjelm~1982, Soderhjelm~1984, Georgakarakos~2002,  Georgakarakos~2009). The third companion induces a long-term (secular) eccentricity modulation, as shown for example in Soderhjelm (1984).

Here, just like in CH2012, we consider a binary formed by two point masses initially in circular orbit.  A third point mass (perturber) moves initially on its own circular orbit, farther away and in the same orbital plane as the other two. Its mass $M_3$ and orbital period $P_3$ are varied across an ensemble of numerical experiments.

\begin{figure*}[t]
\begin{center}
\includegraphics[width=\textwidth]{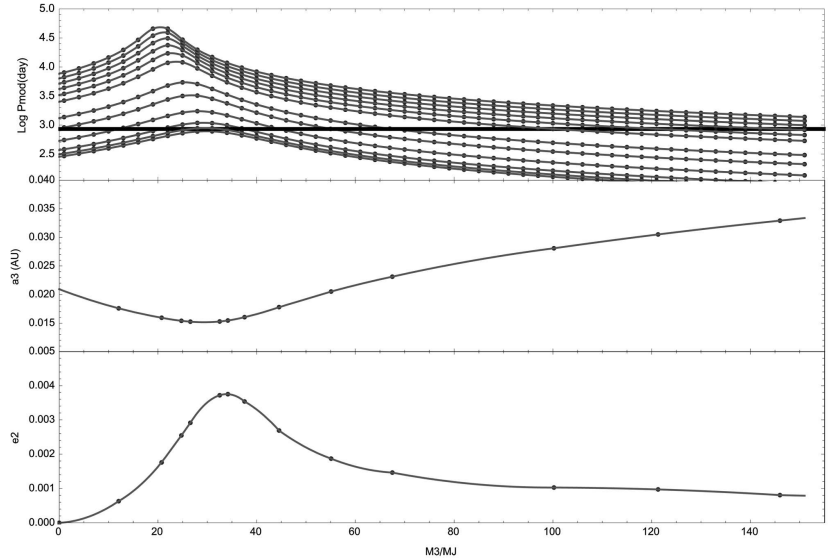}
\caption{The upper panel shows the logarithm of the period of the long--term modulation in the binary eccentricity as a function of the perturber mass (in Jupiter masses). Each curve corresponds to different  $P_3/P_{2}$ ratios taken from 12.5 to 40.8; the values are 12.5, 12.7, 12.9, 13.1, 13.4, 15.6, 19, 22, 30.6, 33.6, 37.2, 40.8, rom bottom to top. The thick horizontal line shows the observed value of the VLPP (857 days). Only solutions that cross this line can explain the VLPP. 
The middle panel shows the perturber mass and semi--major axis combinations that result in a long--term modulation of the binary orbit equal to the VLPP, that is the solutions that cross the black thick line. 
The lower panel shows the amplitude of the binary eccentricity perturbation for those solutions. See text for discussion.}
\label{fig:Pmodm3}
\end{center}
\end{figure*}

The upper panel of Figure\,\ref{fig:Pmodm3} shows the log10 of the resulting  periods of the long-term modulation of the binary eccentricity (vertical axis) as a function of  the mass of the perturber (horizontal axis), for the entire ensemble of our numerical experiments. Each curve corresponds to different  $P_3/P_{2}$ ratios taken from a range of values between 12 to 48; bottom and top curves, respectively.  
The  thick horizontal line corresponds to the VLPP value. For example, the curve with $P_3/P_{2}=12$ does not cross the line and therefore it is a value that cannot explain the VLPP observed. 
For perturbers whose orbital period is smaller than 12 binary periods no solution is possible, since  their respective curves do not reach  the VLPP  value.  For perturbers with  periods longer than that, but shorter than 19 binary periods, two solutions are possible: one at low mass and another at an increasingly larger mass. Finally, perturbers with longer periods than 19 produce only one solution at the  large mass range.

The curve in the middle panel of Figure\,\ref{fig:Pmodm3} shows the perturber's orbit semi-major axis but only for the solutions that could explain the observed VLPP value; i.e solutions that cross the solid line on the upper panel.  
The lower panel  shows the  amplitude  of the eccentricity perturbation for the solutions presented in the middle panel. The most efficient case would be the one in which the VLPP is the predominant effect and the eccentricity pumped in the inner binary is the largest; that is,  the minimum in semi-major axis and the maximum in eccentricity. According to this study, the maximum amplitude is achieved for a system that has a third body with M$_{3}=29$M$_{J}$ and $P_3/P_2=12.7$.

All curves in the upper panel of Figure\,\ref{fig:Pmodm3} gets to its maximum value for smaller values of the mass compared to Fig. 8 upper panel of CH2012 for the same initial conditions.
Therefore, in the middle panel of Figure\,\ref{fig:Pmodm3}, we also obtain smaller values for the masses of the possible third body compared to the middle panel of Fig. 8 of CH2012 for the same initial conditions. Then, the minimum of this curve in this research is obtained when M$_{3}=29$M$_{J}$ and $P_3/P_2=12.7$, while the minimum in CH2012 was obtained when M$_{3}=48$M$_{J}$ and $P_3/P_2=13.4$.

The relative eccentricity amplitudes of these three modulations (inner binary period, third body period and secular VLPP) depend on the mass and size of the orbit of the perturber.
The VLPP modulation becomes the predominant effect in the range of  masses  for a third body of $20M_J < M_3 < 45M_J$. 
 The envelope of  the calculated long-term modulation of the binary eccentricity for our best case is remarkably similar to the waveform of the VLPP.


\subsection{The third body on an eccentric and inclined orbit}

Now we investigate the effect of the eccentricity and inclination of the third body on the outcome of the VLPP.

\begin{figure}
\begin{center}
\includegraphics[width=\columnwidth]{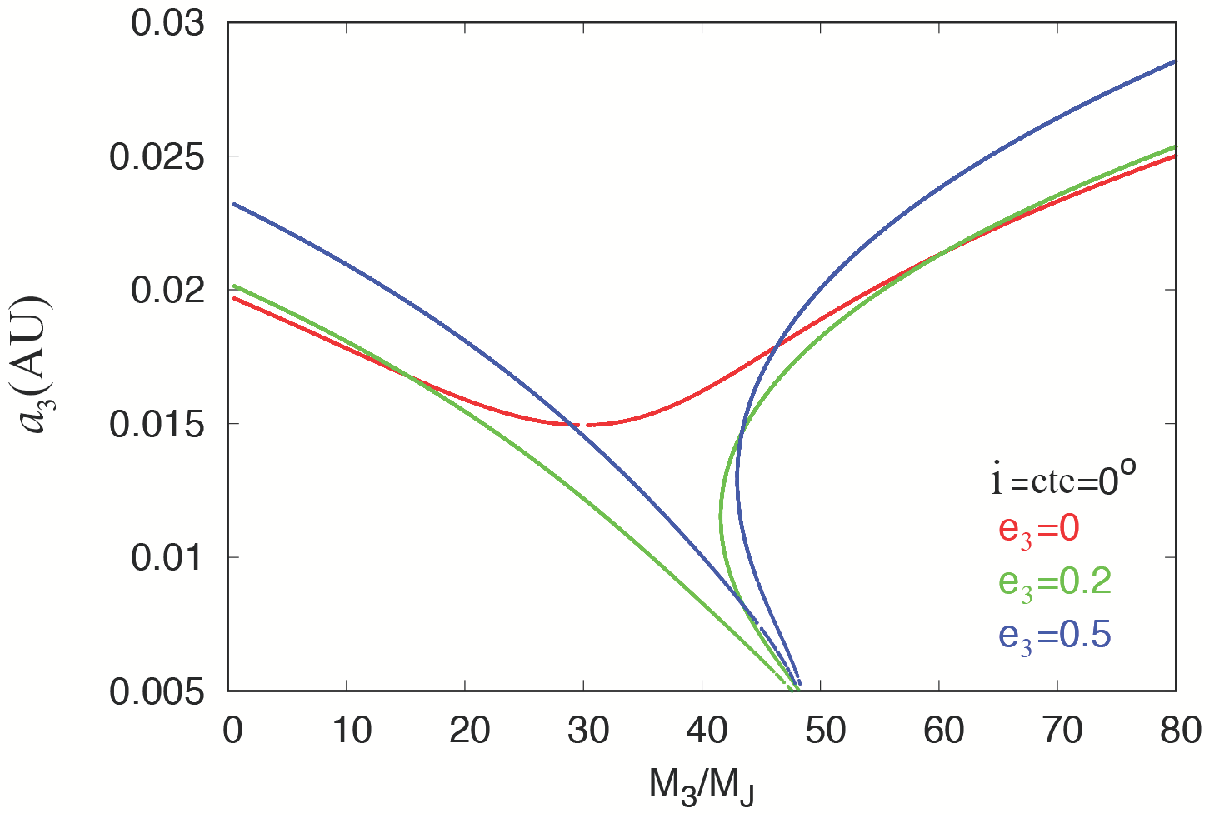}
\includegraphics[width=\columnwidth]{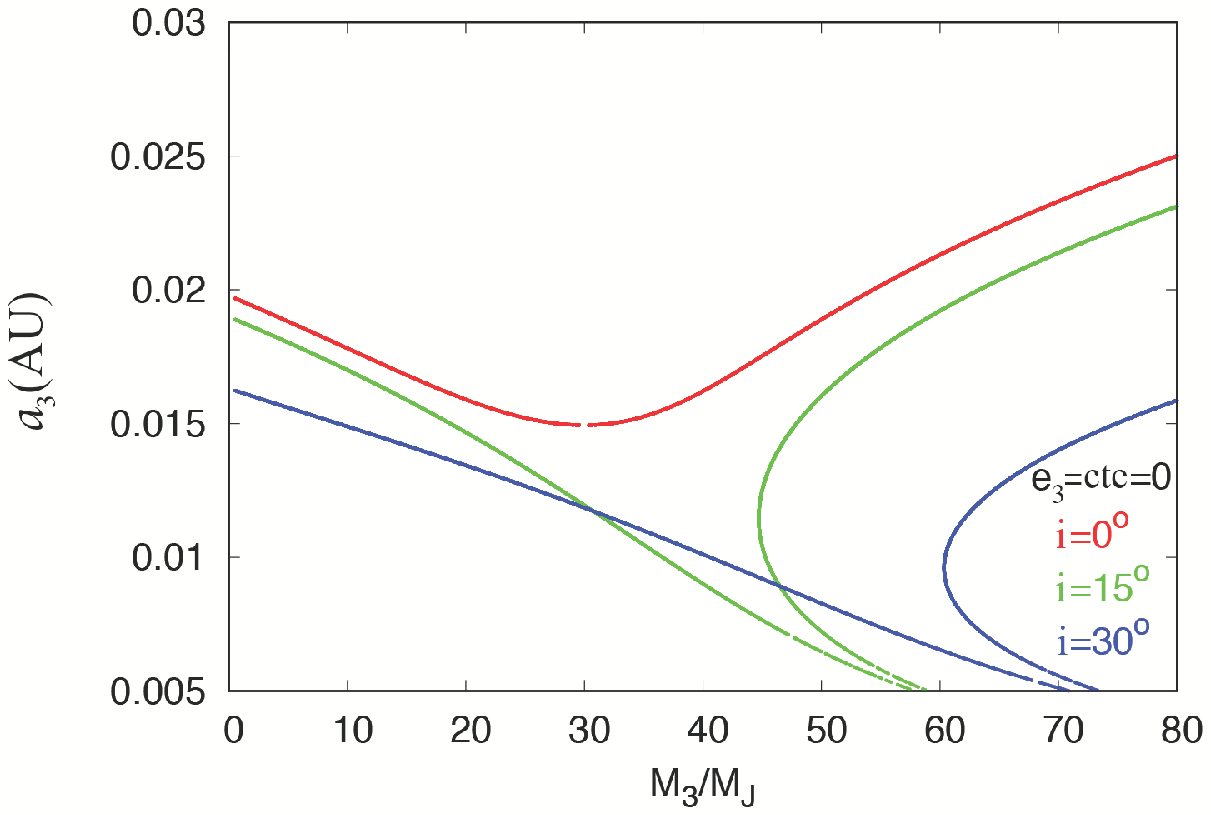}
\caption[]{Perturber mass and semi--major axis combinations that result in a long--term modulation of the binary orbit equal to the VLPP of 857 days. These results were obtained using the analytical formulas described in the text. In the top plot we explore the effect of the eccentricity of the third body, the inclination for all systems remains constant $i=0^{\circ}$. The bottom plot explores the effect of the orbital inclination, the initial eccentricities for all systems is $e_3=0$ .  See text for discussion.} 
\label{fig:PN}
\end{center}
\end{figure}

Figure  \ref{fig:PN} contains two plots for various dynamical scenarios.  It is clear that there is a variety of combinations of masses and semi-major axes of the hypothetical companion that can produce the observed long term variation in the light curve of \fs.  The perturber's eccentricity does not seem to affect
very much when we compare the two analytical solutions for $e_3=0.2$ and $e_3=0.5$.  The low $e_3$ solution seems to be a bit different in the range $M_3/M_J=30-50$.  Similarly, there is some difference among the solutions  as the mutual inclination increases.

The orbital solutions based on our analytical estimates yield a wide range of masses for our hypothetical companion, from sub-Jupiter mass bodies to big brown dwarfs.  However,  
all solutions may not be dynamically stable.  According to the empirical criterion  developed by Holman and Wiegert~(1999), the smallest stable semi-major axis for our unseen companion is 0.0055 AU.  This value is valid for small values of $e_3$, as the criterion was based on simulations of massless particles initially on circular orbits around the binary star. For initially eccentric orbits around the stellar binary the value of the stable semi-major axis may be different.  The same holds when the 
companion has a mass comparable to the secondary;  $M_2=0.079 M_{\odot}$ which is about $83 M_J$, and therefore the empirical criterion of Holman and Wiegert is valid only for masses in the left part of our plots.  
In this case, we can get an idea about the stability limit from Table A1 of Georgakarakos (2013) which provides values for three 
dimensional systems, but only for initially circular orbits however.   Considering the outer mass to cover the range we have in our plots, we find that for coplanar and low inclination systems ($i=20^{\circ}$) the stability limit is around $a_3=0.01$AU.

\subsection{Effect of Post--Newtonian correction}

Here we consider the possible
 dynamical effects that a first order post--Newtonian correction to the binary's orbit may produce the long--term signal we observe in the light curve of the stellar binary. That is the first order general relativity correction in the precessional rate of the longitude of the pericentre.

For the stellar binary under investigation, although its total mass is under one solar mass, 
the small semi-major axis of its orbit makes it an interesting case to consider a post-Newtonian correction.  The consequence of including this effect to our orbit results in the precession of the pericentre at the following rate (Georgakarakos,~\&~Eggl~2015, Naoz~et~al.~2015):
\begin{equation}
\dot{\varpi}=\frac{3 {G}^{\frac{3}{2}}(M_1+M_2)^{\frac{3}{2}}}{c^2a^{\frac{5}{2}}_1(1-e^2_1)},
\end{equation}
where $G$ is the gravitational constant and $c$ is the speed of light in vacuum.

Based on the precession rate given by the above equation, the period of the pericentre circulation for our system is 6812 days (18.65 yrs).
Since this number is much larger than the 857 day signal we observe in the light curve of the system, we conclude that general relativity (GR) by itself cannot explain it.

\subsection{Estimation of the effect of the third body on the mass transfer rate and brightness of the system}

Now that we have established that a third body can explain the VLPP observed, we estimate how the modulation of the inner binary due to the secular perturbation of the third body affects the mass transfer and then the brightness of the system.

The results of our numerical integrations for the third body on a close near-circular and co--planar orbit,  the most efficient solution is used for all calculations in this subsection; that is M$_{3}=29$M$_{J}$,  $P_3/P_2=12.7$,  $P_3=18.16$ h. In order to estimate the mass loss of the secondary we make use of the concept of RL(2). Since calculating the volume of the Roche lobe is difficult, we can define an equivalent radius of the Roche Lobe as the radius, RL(2), of a sphere with the same volume as that the Roche lobe. This radius RL(2) has been widely studied for different mass ratios ($q={M_1 / M_2}$) between the primary and the secondary. Equation 2 by Eggleton (1983) is widely used since is valid in a wide range of mass ratios (valid for $0<q<\infty$) and accurate to better than $1\%$. That equation assumes that the orbit is circular and that the semi--major axis is constant.

\begin{figure}
\begin{center}
\includegraphics[width=\columnwidth]{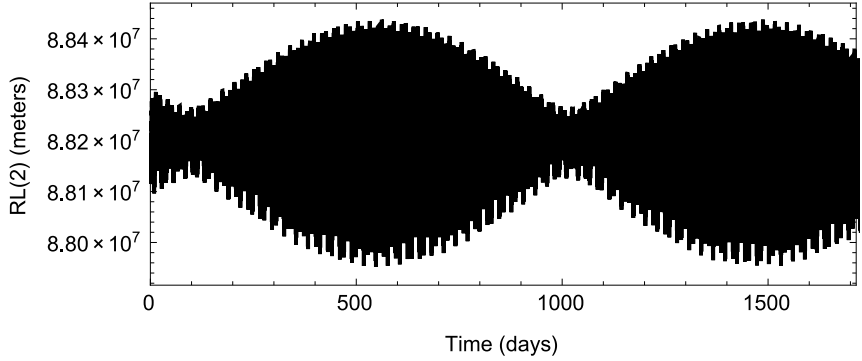}
\caption[]{ Location of RL(2) as a function of the time, RL(2) is the radius of the sphere with volume equal to that of the Roche lobe of the system.  See text for discussion.} 
\label{fig:RL2}
\end{center}
\end{figure}

Sepinsky et al. 2007 studied the definition of RL(2) for eccentric binaries finding the following generalisation:
\begin{equation}
RL(2)=r_{12}(t){{0.49 q^{2/3}} \over {0.6 q^{2/3} + \ln{(1+q^{1/3})}}},
\end{equation}
where $r_{12}$ is the distance between the two stars at a given time. Since we have that distance from our integration of the most efficient case, we can plot RL(2) as a function of time as in shown in Figure\,\ref{fig:RL2}. 

We can calculate the maximum $RL(2)_{max}=8.844\times 10^ {7}$m and $RL(2)_{min}=8.796\times 10^ {7}$m, from here in principle we can estimate the mass transfer rate $\dot{M} (2)$ and from here the luminosity of the Cataclysmic Variable.

We proceed as follows. First we assume that the secondary is a polytrope of index 3/2 (we assume certain shape of the Roche Lobe). Also that the density around L1 point is given by Eq. 2.11 of Warner (1995), $\rho_L1=\rho_{0} e^{- (\Delta R / H^{\prime})^2}$; where $\rho_{0}$ is the density of the isothermal atmosphere, and $H^{\prime}$ is a scale height given by Lubow \& Shu (1975).

We can estimate the mass transfer rate using the Eq. 2.12 of Warner (1995),

\begin{equation}
\dot{M}(2)= - C {M(2) \over P_{12}}   \Bigg({\Delta R \over {R(2)} } \Bigg)^{3},
\end{equation}

\noindent where $C$ is a dimensional constant $\approx 10-20$ and $\Delta R$ is the amount by which the secondary overfills its Roche Lobe: $\Delta R=R(2)-R_{L}(2)$.
The $R(2)$ distance needs to be calculated carefully since the equation for $\dot{M}(2)$ is very sensitive to the amount of overfill. We decided to adjust the $R(2)$ to obtain the $\dot{M}(2)$ that we report here in Table \ref{tab:initial}; the logarithm of the secondary star mass loss rate of $-10.25  \big({M_{\odot} \over  yr}\big)$. Since $R_{L}(2)$ is a function of the time we use the mean value of $R_{L}(2)_{mean}=8.821 \times 10^{7}$ m for the $R_{L}(2)$ value. Hence we obtain the value $R(2)=8.820 \times 10^7 $ m.

Therefore, we can calculate the maximum and minimum of the mass transfer rate by using the values of $RL(2)_{max}$ and $RL(2)_{min}$. We obtained $\dot{M}(2)_{max}=7.1 \times 10^{18}$ kg/s and $\dot{M}(2)_{min}=5.8\times 10^{18}$ kg/s.

We make an estimation on the luminosity due to the accretion (Warner 1995). First, calculate the luminosity due to the so called hot spot (the place where the stream of stellar mass crosses the L1 point and collides with the disk): 

\begin{equation}
L(SP) \approx {G M(1) \dot{M}(2) \over r_{d}},
\end{equation}

\noindent where $L(SP)$ is the luminosity due to the hot spot, the radius of the disk is typically $r_{d} \approx 0.40\times a_{12}$, where $a_{12}$ is the semi--major axis of the inner binary,  both given in Table \ref{tab:initial}.
Applying this equation to our extreme values on RL(2) we obtain: $L(SP)_{max} \approx 4.2 \times 10^{30}$ Watts and $L(SP)_{min} \approx 3.2 \times 10^{30}$ Watts.
The luminosity of the accretion disk ,using Eq. 2.22a of Warner (1995), is:

\begin{equation}
L(d)\approx {1 \over 2} {G M(1) \dot{M}(2) \over R_{1}},
\end{equation}

Using this equation for $L(d)$ we can obtain the extreme values of $L(d)_{max} \approx 4.8 \times 10^{31}$ Watts and $L(d)_{min} \approx 3.6 \times 10^{31}$ Watts. The total luminosity for each extreme is obtained by adding the estimated luminosity of the hot spot plus the luminosity of the disk, obtaining: $L(d)_{T_{max}} \approx 5.2 \times 10^{31}$Watts  and $L(d)_{T_{min}} \approx 4.0 \times 10^{31}$ Watts.

We can calculate the bolometric magnitude using $M_{bol}=-2.5 \log \big({L \over L_{0}}\big)$, with $L_{0}=3.0128 \times 10^{28}$ Watts. For the extreme values we obtained  $M_{B_{max}}=-8.09$ and $M_{B{min}}=-7.79$, giving us a magnitude difference of $\Delta M_{B} = 0.29$.

The observed change on magnitude at quiescence is $\approx 0.4$ magnitudes when using a sinusoidal best fit, as shown in Figure\,\ref{fig2new}, but the data points are quite disperse, so we also calculated the difference between the maximum and minimum magnitude of the observed data finding 1.1 magnitude.
The most efficient parameters model give us an expected change of magnitude of $\approx 0.29$. We remind the reader that the later was an order of magnitude estimate with simplifications, assumptions and estimations.


\section{Alternative Scenarios for the VLPP}

One possible explanation to the VLPP is the cyclic magnetic variation, analogue to the Solar cycles, in the secondary star which may lead to mass transfer variations. Long variations have been observed in CVs as mentioned in Richman et al. (1994), where they concluded that this explanation is plausible. But they found that these cycles did not show any strict periodicity and are decades long. In Table 3 in Mascare\~no et al. (2016), the magnetic cycle of medium to late M stars is calculated and found to be 7.1 years for a sample of this type of stars.

As pointed out at the end of Section 3.1 in this research, the secondary star on FS-Aur is expected to be a very late M star, their internal structure not being the same as their normal main sequence star counterpart with the same mass. Stars with $M\approx 0.4M_{\odot}$ become fully convective as the mass decreases, the density increases and the internal temperature decreases, leading to the partial degeneracy of the core. Approaching the minimum hydrogen-burning mass of $0.08 M_{\odot}$, the increased electron degeneracy induces structural changes on the secondary. Making the star magnetic but with very few spots.

Works of Bianchi (1992) and Hessman et al. (2000) found evidence of a possible relation between mass accretion variations and solar cycle type phenomena. The evidence showed variations on the timescales of decades on overall system brightness and gave theoretical support for star-spots migrating to the L1 region (Howell 2004). This migration would help to correlate the star-spot to the changes in the position of the L1 point due to a possible third body.

Nevertheless, the magnetic cycles in very late M stars have not been studied in detail for secondaries in CVs. and we recognise this mechanism as a strong alternative to the mechanism proposed here.

\section{Summary and Final Comments}
\label{Sec:final}

We confirm the presence of VLPP with a refined period of 857 days based on 5 more years (20 years total) of observations for \fs.  This result also helps in confirming the authenticity of this signal.

We also revisited the triple CV hypothesis in which a massive planet, or a substellar object, pumps eccentricity into the inner binary orbit by secular perturbations. New parameters of mass, radius and temperature for the binary members of the CV \fs\/ Knigge et al. 2011 were calculated, and we used these to recalculate the most efficient parameters for the third body as defined earlier. The most efficient combination that explains the 857 day period is a third body with M$_{3}=29$M$_{J}$ and $P_3/P_2=12.7$ ($P_3=18.16$ h). This new value is 1.7 times less massive than our previous estimation and is well within the limits of planetary mass. For example, the planet HD 169142b has a similar mass Fedele et al. 2017. All these numerical calculations were made for a third body in an initial circular and planar orbit as in CH2012.

We also explored more complications to the model  to study the secular perturbations of systems with eccentric and inclined orbits, using previous analytical results (Georgakarakos 2002, 2003, 2004, 2006, 2009).
 We found that as the eccentricity increases the most efficient candidate third body has a larger mass: M$_{3}=47$M$_{J}$ for an eccentricity of 0.2, and M$_{3}=48$M$_{J}$ for an eccentricity of 0.5 of the third body.

 When the mutual inclination is explored the most efficient candidate for the third body has larger mass: if the inclination is 15$^\circ$ the expected most efficient mass is about M$_{3}=58$M$_{J}$, but when the inclination is 30$^\circ$ the expected most efficient mass now is about M$_{3}=72$M$_{J}$.

We considered other dynamical effects that might produce this VLPP, such as the first order post--Newtonian correction. We found that for \fs\ the period of the pericentre circulation is 6812 day (18.65 yrs), that is much larger than the 857 day period observed.

We calculated a first order estimation of the effect of the secular period due of the third object on the mass transfer rate and then on the brightness of the system; a change of magnitude of the order of only $\Delta M_{bol} = 0.29$ was obtained. Even when this change is not the 0.4--magnitude observed, is quite close for an order of magnitude calculation. It also gave us insights on how sensitive is the system to even smallest changes in the parameters to calculate $\dot{M}(2)$, to show that we changed the distance $R(2)$ by less than $0.01\%$ and we obtained the 0.4--magnitude observed. The $R(2)$ adjustment was based on the value of $\dot{M}(2)$ that appears in Table \ref{tab:initial} taken from Knigge et al. 2011 and that value was calculated using statistics. The change in magnitude of \fs~are may be a mechanism to explain the VLPP observed.
 
We examined alternative scenarios for the VLPP. A possible explanation by a solar type magnetic cycle of the secondary cannot be ruled out for the VLPP, since the VLPP is only 2.346 years and most of the cyclic type magnetic periods in mid to late M stars are of the order of decades.However there are no studies for the magnetic cyles of very late M stars in CVs to asses further this hypothesis, then making this alternative a plausible one.

In summary, we find (a) that the new extended data confirms that there is a VLPP, but with a new value of 857 days,
(b) These new data is consistent with a triple-system for \fs, (c) that combining such data with new initial conditions yield a reduction (from  M$_{3}=50$M$_{J}$ to  29M$_{J}$) in the mass estimate for the third-body most efficient candidate in \fs., (d) an order of magnitude estimation for the mass transfer rate and the brightness of the system has been done, with the initial conditions used here, lead to a change on magnitude  of 0.3. This value was 25\% times smaller than the observed but we found that changes of less than $0.01 \%$ in the R(2) parameter increases the change in magnitude to the observed one.

\section*{Acknowledgements}
 We would like to thank all the amateur  observers who do a great hard job in collecting professional grade data with persistence. We are particularly indebted to Joe Patterson, who guides the amateur community engaged in CV monitoring and who made possible the dense observational coverage of \fs. We acknowledge with thanks the variable star observations from the AAVSO International Database contributed by observers worldwide and used in this research.
CC acknowledges UANL PAICYT grant. We appreciate the comments, suggestions and corrections by the anonymous referee, which helped us to greatly improve the quality and content of this research.


\end{document}